\documentclass[journal]{IEEEtran}

\usepackage{bibentry}  
\usepackage{xcolor,soul,framed, balance} 

\colorlet{shadecolor}{yellow}
\usepackage[pdftex]{graphicx}
\graphicspath{{../pdf/}{../jpeg/}}
\DeclareGraphicsExtensions{.pdf,.jpeg,.png}
\usepackage{dingbat}
\usepackage[cmex10]{amsmath}
\usepackage{array}
\usepackage{mdwmath}
\usepackage{mdwtab}
\usepackage{eqparbox}
\usepackage{url}
\usepackage{siunitx}
\usepackage{booktabs}
\usepackage{pifont}
\usepackage{multirow}
\usepackage{amsmath}
\usepackage{mathtools}
\usepackage{breqn}
\usepackage{graphicx}
\usepackage{array}
\usepackage{setspace} 

\begin{document}
\bstctlcite{IEEEexample:BSTcontrol}
    \title{Key Management Systems for Smart Grid Advanced Metering Infrastructure: A Survey}
 \author{Amrita~Ghosal\thanks{Amrita Ghosal and Mauro Conti are with Department of Mathematics, University of Padua, Via Trieste 63, Padova, Italy, 35121 (e-mail: amrita.ghosal@math.unipd.it, conti@math.unipd.it).} and 
        Mauro~Conti,~\IEEEmembership{Senior Member,~IEEE}}

%
%


\maketitle
\begin{abstract}
Smart Grids are evolving as the next generation power systems that involve changes in the traditional ways of generation, transmission and distribution of power. Advanced Metering Infrastructure (AMI) is one of the key components in smart grids. An AMI comprises of systems and networks, that collects and analyzes data received from smart meters. In addition, AMI also provides intelligent management of various power-related applications and services based on the data collected from smart meters. Thus, AMI plays a significant role in the smooth functioning of smart grids. 
\par AMI is a privileged target for security attacks as it is made up of systems that are highly vulnerable to such attacks. Providing security to AMI is necessary as adversaries can cause potential damage against infrastructures and privacy in smart grid. One of the most effective and challenging topic's identified, is the Key Management System (KMS), for sustaining the security concerns in AMI. Therefore, KMS seeks to be a promising research area for future development of AMI. 
This survey work highlights the key security issues of advanced metering infrastructures and focuses on how key management techniques can be utilized for safeguarding AMI. First of all, we explore the main features of advanced metering infrastructures and identify the relationship between smart grid and AMI. Then, we introduce the security issues and challenges of AMI. We also provide a classification of the existing works in literature that deal with secure key management system in AMI. Finally, we identify possible future research directions of KMS in AMI.
\end{abstract}

\section{Introduction}

\IEEEPARstart{S}{mart} Grids reorganized the traditional concept and operation of electrical grids by using information technology~\cite{Wu11},~\cite{Pindoriya2013}. Also, maximum utilization of information technology is done in smart grids for achieving system efficiency and reliability~\cite{Wang13}. In addition to power generation and transmission utilities, smart grids consist of appliances, meters, sensing devices, information gateways that operate in near real time~\cite{Khasawneh17}. The meters perform the task of collecting the energy consumption of appliances, broadcasting energy loss/restoration information and reporting the pricing information to customers. The sensing devices are deployed to monitor the system performance and detect any operation malfunction. Upon detection of any failure, the sensing devices transmit control messages to the control center. Due to the fact that smart meters are located far from the utility, intermediate devices are needed to route the smart meters data to the utility. Gateways (sometimes called concentrators) collect smart meter's data and send it to the utility over Wide Area Network (WAN) connection. Control information is also propagated through the gateways to the smart meters. 
\par
\begin{figure}[h]
 \centering 
 \includegraphics[width=\columnwidth]{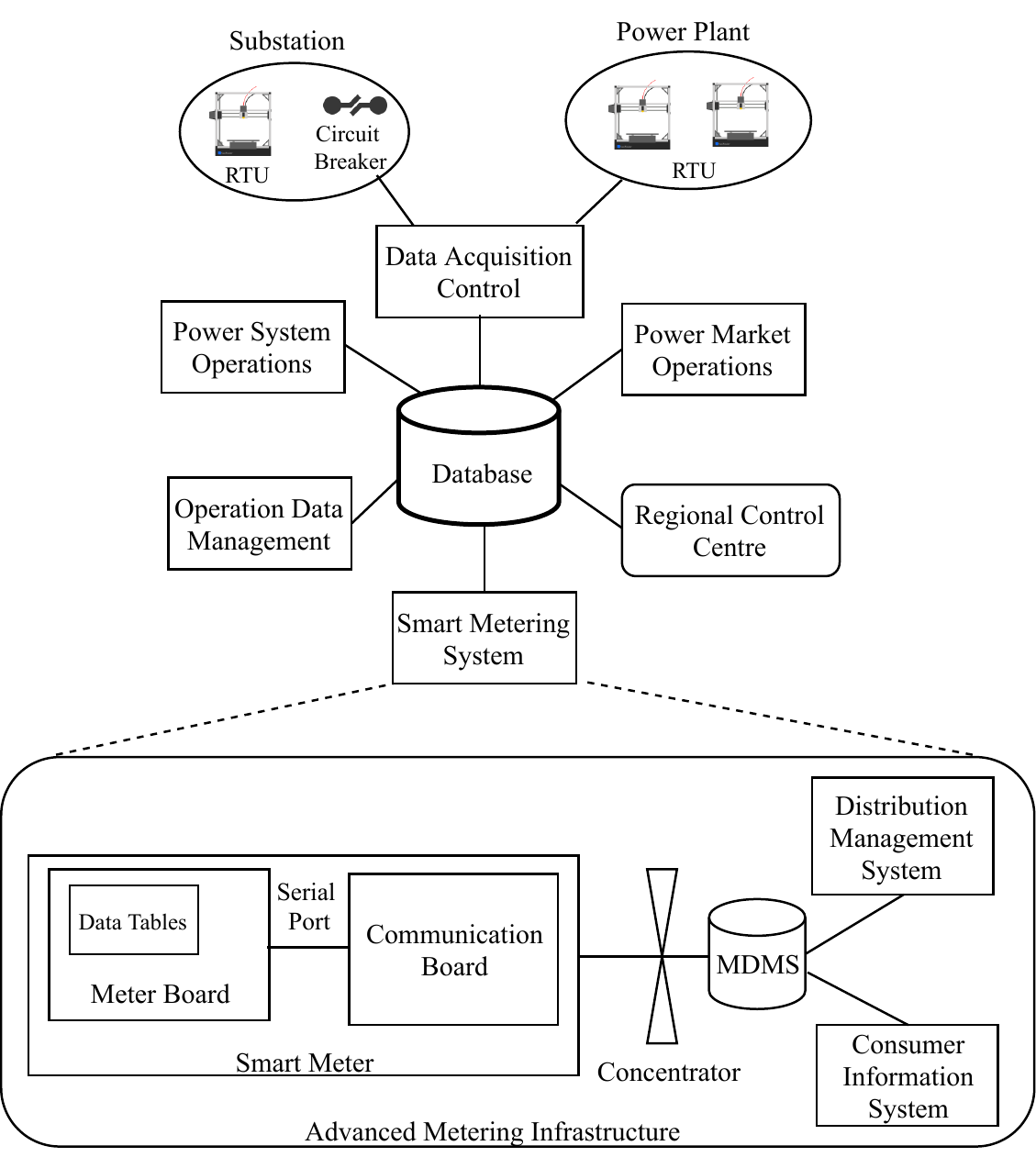}
 \caption{A Smart Grid Communication System.}
 \label{fig:Fig_1}
\end{figure}

 The architecture of smart grid necessitates the smart meters, sensing devices, gateways and the control centers to exist in the path between the customers and the power providers for realizing the two-way communication~\cite{Yan12},~\cite{Fang12}. A number of factors are involved in designing of a smart grid, but mainly, coordination among three fields of communication, control and optimization is highly essential. Ideally, a smart grid design should address reliability, adaptability and prediction issues. Also, the designing should address the challenges for load handling and demand adjustment, incorporation of advanced services, flexibility and sustainability, end to end control capability, power and service quality, cost and asset optimization, security, performance, self-healing and restoration. In general, a smart grid communication system is composed of a horizontal integration of one or more regional control centers, where each center is responsible for supervising the operation of multiple power plants and substations~\cite{Yan12}. Figure~\ref{fig:Fig_1} shows the structure of a smart grid communication system that does data collection and control of electricity delivery. The smart grid consists of components such as, regional control center, substation, smart metering system and power plants. A regional control center typically supports metering system, operation data management, power market operations, power system operation and data acquisition control. Substations consist of Remote Terminal Units (RTUs)~\cite{Deng17},~\cite{Choi10} circuit breaker, human machine interfaces, communication devices (switches, hubs, and routers), log servers, data concentrators, and a protocol gateway. Intelligent electronic devices are field devices, that include an array of instrument transducers, tap changers, circuit re-closers, phase measuring units, and protection relays~\cite{Yan12}.

Advanced Metering Infrastructure (AMI)~\cite{Erol-Kantarci15}, that also goes by the name ``smart metering"~\cite {YYan13} is a critical component, that necessitates the realization of the vision of ``Smart Grid". In general, AMI consists of the smart meters~\cite{Yan11},~\cite{Odelu16}, concentrators and the Meter Data Management System (MDMS)~\cite{US_report2}. The smart meters contains the meter board and the communication board connected through a serial port. The meter board contains a set of tables storing various information including keys and passwords used for secure communication and privilege levels. It also performs power consumption measurements. The communication board is responsible for communications with the outside nodes such as collectors, other smart meters, or home appliances, and for performing any required computation. Using an interrupt based mechanism, the communication board fetches data and other necessary information such as keys from the meter board whenever it needs to send data to the utility. The MDMS is a key component in AMI, that is basically a database responsible for long term data storage and management of the huge amount of usage data and events.
\par AMI also facilitates two-way communication~\cite{Mohassel14} between meter and distribution system operator. The two-way communication enables several services for the distribution system operator that were difficult or impossible to implement without smart metering. For example, power outage is detected faster by system operator and without interaction with the customer. Another service provided by smart metering is reporting the quality of power delivery (voltage, frequency). Smart metering also enables detailed monitoring of power flows within the distribution system that was previously available only at the substation level. Monitoring of power flows is important as it enables energy suppliers to react quickly on variations in consumption levels. The power flow monitoring information is also useful for real time pricing, that is handled by one technology in smart grid known as Demand Side Management (DSM)~\cite{Kabalci16}. If prices for energy are variable and react on current power flow information, real time pricing comes into the picture. This is one feature of demand side management, where the energy supplier influences the consumption of energy directly and immediately~\cite{Finster14}.

The use of wireless communication in AMI leads to security issues in such systems. There are several security issues with regard to AMI that needs attention, ranging from the consumer level to the generation as well as the producer level. The adversary can launch an attack by sending false signals to meters that may lead to power outage in a particular area as well as disturb the demand generation model. The adversary can also make use of the study of the utilization pattern of the consumers for devising new forms of attacks.
\par Similar to other existing systems, AMI too needs to adhere with the requirements of the security primitives of confidentiality, integrity, availability~\cite{Gungor11} and non-repudiation.
\begin{itemize}
\item \textbf{Confidentially} is preserved in AMI by ensuring that the energy consumption pattern of consumers are not revealed to unauthorized entities.  
\item \textbf{Integrity} is maintained in the system through detection of illegal data alternation. 
\item \textbf{Availability} requires the accessibility of data by an authorized user on demand. If the required data is not found at the time of need, the system violates the availability aspect of the security requirement of the system. Any natural or intentional incidents (such as hacking) must not hamper the system from operating correctly. For example, if the hacker wants to jam the network, the system must comply with the availability aspect. 
\item \textbf{Accountability} (non-repudiation) means an action that cannot be denied, i.e., the entities used in receiving or transmitting data must not deny it. In the AMI network, accountability ensures a timely response to the command and control, and integrity of billing profile, etc. 
\end {itemize}
\par To meet the security requirements stated above, cryptographic countermeasures must be deployed. But cryptographic mechanisms for AMI also require an efficient key management~\cite{Hasan2015},~\cite{Seo2013}. Inadequate key management may result in possible key disclosure to attackers, and even endanger the secure communications in AMI~\cite{Benmalek2016}. Therefore, key management is a critical process and can be used as a defensive mechanism against threats and vulnerabilities~\cite{Das12}. Generally, the security requirements of AMI include confidentiality, integrity, and availability~\cite{Kamto11},~\cite{Rabieh17}. Before AMI deployment, the confidentiality for customer privacy and customer behavior, as well as message authentication for meter reading, Demand Response (DR)~\cite{Deng15}, and load control messages, are the most important security requirements to be provided. Confidentiality and integrity are ensured by encryption and authentication protocols, that depend heavily on the security of cryptography keys. 
\par To ensure the security of cryptographic keys, the key management for large amounts of devices in AMI systems is very important. Recently, several studies were conducted related to the Key Management System (KMS)~\cite{Mohammadali18} of AMI, a vivid description of the same is provided in Section III. Existing surveys on smart grids discussed topics on architectures, applications, communication, smart metering and cyber security. In contrast, this survey work deals with key management systems for smart grid in AMI, a very critical area where very less attention has been paid to. This survey showcases the importance of AMI in smart grids and also focuses on the key management system that plays a defensive role in AMI against threats. The novelty of this survey is in the classification of the existing works and outlining of future research in key management system of AMI. The main contributions of this survey are:
 
\begin{itemize}
\item We discuss the significance of AMI in smart grids as well and provide a classification for the current state-of-the-art works in key management systems of AMI.
\item We provide a comprehensive discussion on the system structure of AMI, followed by the issues and challenges faced by AMI.  
\item We classify the existing works in literature, based on their commonality in approaches. 
\item We present a comparative study for the existing works in key management system schemes in AMI, considering communication, computation and storage overheads as performance metrics. The metrics chosen for the comparative analysis reflects the efficiency of the corresponding scheme.
\item We present promising directions for future research in smart grids, particularly in the area of advanced metering infrastructure where further research is required.
\end{itemize}

The rest of this paper is organized as follows. In Section II, description of AMI system features, followed by security challenges and the role of key management system in AMI are carried out. The classified approaches for existing key management systems in AMI are discussed in Section III. In Section IV, we present a comparative study for the current state-of-the-art key management schemes proposed for smart grid AMI. Finally, we identify the future directions and provide conclusion in Sections V and VI, respectively.

\section{Background}
This section provides an insight into the Advanced Metering Infrastructure including the system features (Section II.A) together with the various security challenges (Section II.B) involved in such a system. We discuss the role of key management systems in AMI in Section II.C. 
\subsection{AMI System Features}
\subsubsection{Introduction}
The Advanced Metering Infrastructure is referred to as the system responsible for collecting, measuring and analyzing energy usage from networks that are connected to next generation electricity meters, or popularly known as smart meters. An AMI comprises of software, hardware, communication networks, customer associated systems and a MDMS~\cite{George16},~\cite{Zhou12},~\cite{Sharma15}. Similar to other systems, the smart grid AMI also faces security threats from both inside and outside the system. The utility services of AMI provided to customers should ensure that the adapted security technologies are not vulnerable to adversarial activities. Deploying AMI in smart grid should guarantee confidentiality of user privacy along with authentication for meter readings and control messages.

\begin{figure}[h]
 \centering 
 \includegraphics[height=70mm, width=70mm]{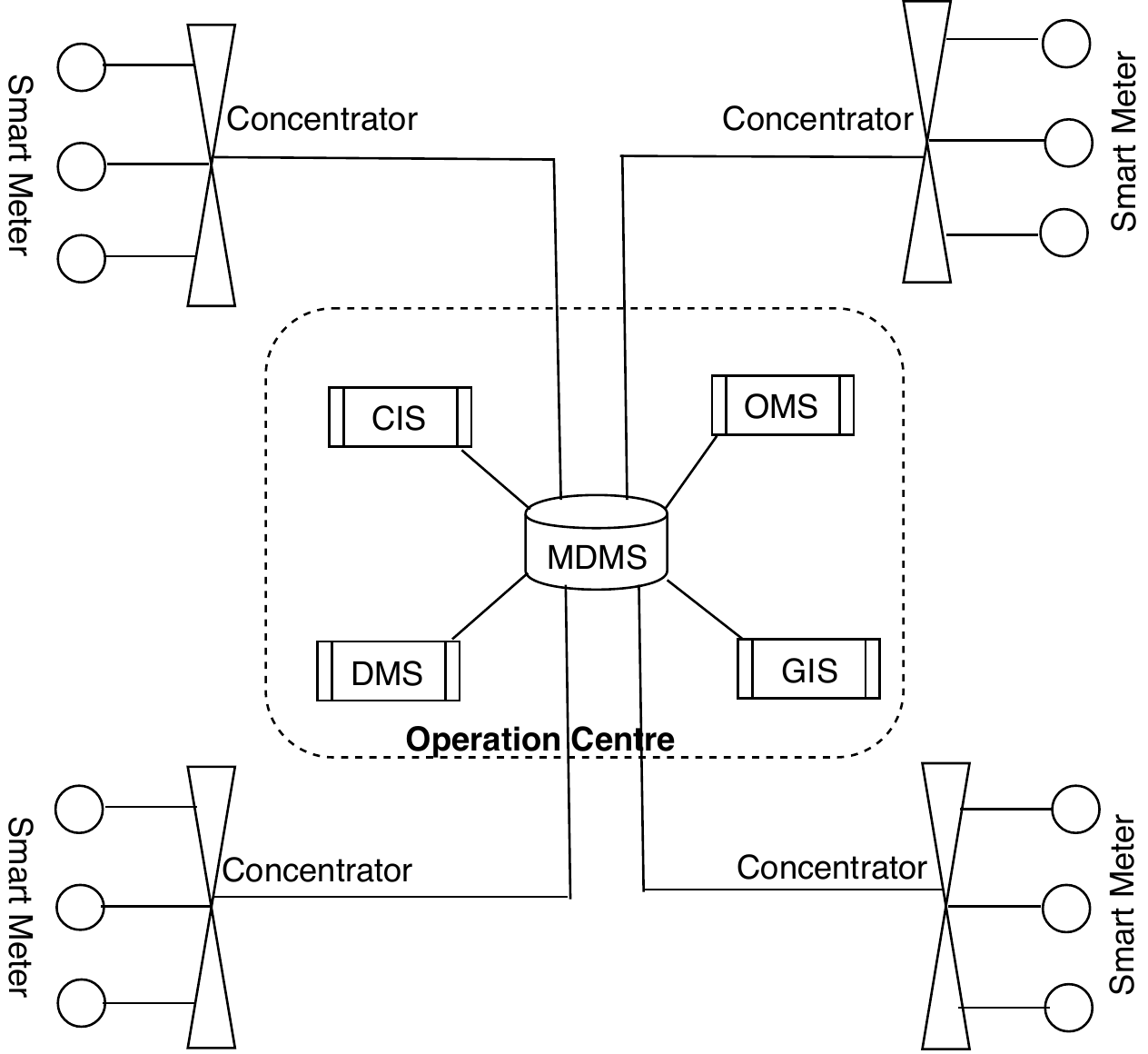}
 \caption{AMI Communication Architecture.}
 \label{fig:Fig_2}
\end{figure}

A traditional AMI communication architecture shown in Fig.~\ref{fig:Fig_2}, has a centralized MDMS surrounded by the main operation and management services. The MDMS is composed of analytical tools that enable the interaction with operation and management systems including Outage Management System (OMS), Geographic Information System (GIS), Consumer Information System (CIS) that manage the utility billing and customer information, and Distribution Management System (DMS) that provides power quality management and load forecasting based on meter data.
\subsubsection{AMI System Structure}
\begin{figure}[h]
 \centering 
 \includegraphics[height=75mm, width=85mm]{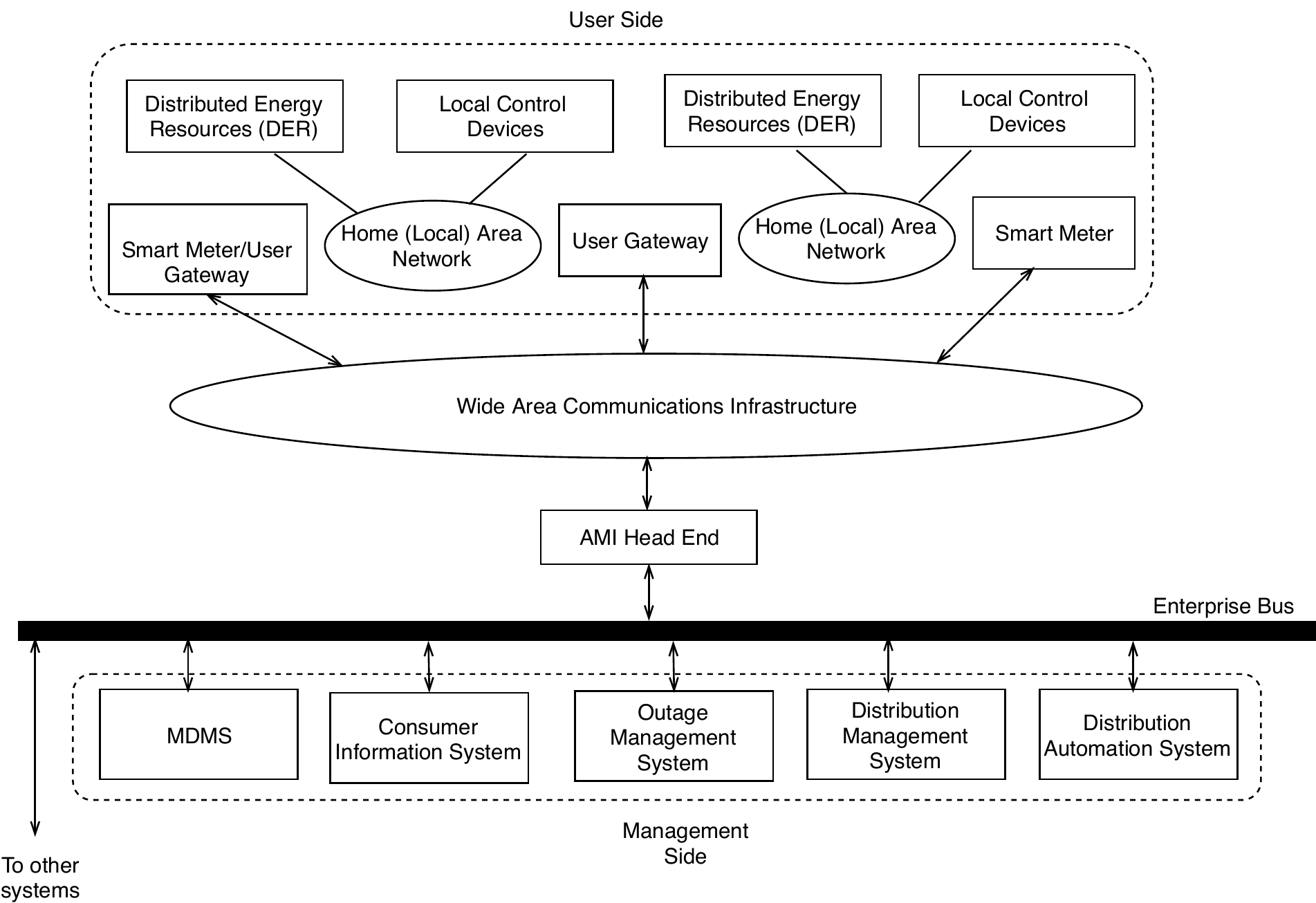}
 \caption{AMI System Structure.}
 \label{fig:sample}
\end{figure}

An AMI is perceived as an infrastructure that integrates several technologies for achieving certain specific objectives. Figure~\ref{fig:sample} describes the components of the AMI system structure that are briefly discussed below: 

\textit{Smart Meters (SMs):} Electrical meters that are responsible for providing two-way communication~\cite{Saputro11}, automated meter data collection, outage management and also allow dynamic pricing.

\textit{Distributed Energy Resources (DERs):} Small scale renewable electricity generation systems that are used for family and energy storage~\cite{Colak16}.

\textit{Gateways (GWs):} Perform the task of implementing protocol conversion and communications between two heterogeneous networks, such as in-home network and wide area network.

\textit{Wide Area Communication Infrastructure:} Provides bidirectional communication between costumers domain and the utility system. Various architectures and medias are used like power line communication system, cellular networks, or IP-based networks~\cite{Sauter11}.

\textit{Meter Data Management System (MDMS):} Acts as database system for storing, managing, and analyzing metering data for proposing dynamic pricing, better customer service, demand response and energy consumption management purposes~\cite{Wan14}.

\textit{Demand Response (DR) Program:} is generally an agreement between the utility and its customers where the customer is ensured of reduced tariffs or discounts in the end-of-month electricity bill, provided that he agrees to reduce his electricity consumption in response to signals received by the grid. The basic concept of the demand response program is that if every customer conserves a little, there will be enough power for everyone~\cite{Kabalci16}. Currently there are different Demand Response programs, each of which has its own policies in terms of rewards, penalties, consumer notification policies and consumer cooperation bases. The benefits of all these demand programs are same irrespective of their particular characteristics.

\par From communication perspective, the AMI comprises of the following networks:

\textit{Home Area Network (HAN):} This type of network connects smart meters and smart devices within home premises~\cite{Wang11},~\cite{Jokar16}. It also provides low cost monitoring and control of the electric devices to reduce energy consumption. Both low range wired and wireless technologies are used for building such networks, though wireless technologies are more dominant. These technologies include 2.4 GHz WiFi, 802.11 wireless networking protocol, ZigBee and HomePlug~\cite{US_report1}. The smart meters are made up of many sensors and data sources. Lightweight security mechanisms are needed for the sensors used in the smart meters as they are generally resource constrained \cite{Taneja2013}.

\textit{Wide Area Network (WAN):} This network performs the task of connecting an AMI end in the local utility network and a data concentrator~\cite{Kuzlu17},~\cite{Chim15}. Data concentrator collects data from a group of SMs over a network and sends this bulk data to the headend. AMI headend manages the information exchanges between external systems, such as MDMS and AMI network.

\textit{Neighborhood Area Network (NAN):} It is mainly the combination of a number of HANs, where all types of necessary information such as power consumption data, appliances control and security alarms are transmitted to achieve energy management~\cite{Kuzlu17}.

\textit{Smart Meter Gateway (SMGW):} It is the central communication component of smart grid infrastructure, that connects a WAN with a network of devices of one or more smart meters. SMGW maintains communication between the consumer with his consuming and generating devices and are secure from physical attacks~\cite{Lunkeit013}. SMGW has a security module that takes care of basic authentication and aggregation of messages sent by various meters on their way to the control center~\cite{Sikora2013}. 
\subsection{AMI Security Challenges}

With the rapid growth in the development of smart grids especially in the context of smart cities, have led to further advancement in technologies such as AMI used in such systems. Security challenges in AMI can in general result from three different aspects: privacy preservation of end users, system resilience against cyber attacks and power theft. In this section, we discuss the context that gives these challenges urgency as well as the technical challenges that need to be addressed by smart grid
communication infrastructures. 

\subsubsection{Privacy Preservation of End Users}

With the increase in the number of smart meters, security issues related with smart grid and AMI continue to scale up considerably, both from internal as well as external part of the system. The consumers' life style can be exposed from the information obtained from the consumers' electric energy consumption, resulting in a critical situation. Examples of leakage of critical information can be in the form of, number of people living in the house, duration of occupancy, type of appliances, security and alarming systems, special conditions such as medical emergencies and so on. Many studies have shown how vital information is obtainable by use of consumer profiling~\cite{Murrill12},~\cite{Molina-Markham10}. Some work, such as~\cite{Kalogridis10} introduced techniques for improving the privacy protection of consumers data. The method used in~\cite{Kalogridis10} mainly reshapes the overall pattern of data such that it is impossible to differentiate between load patterns and signatures. This is done by using three techniques of hiding, smoothing and mystifying consumption using combination of grid and storage/battery as power source. \par For expansion of AMI, consumers satisfaction is very much important. If consumers experience poor service or power quality due to external system manipulation by unauthorized parties or hackers, then, they may resist the implementation of AMI. The price signal and commands obtained at the consumer end are also probable areas for cyber and physical attacks for the purpose of infrastructure damage or power theft. Also, long distance transmission of data as well as storing the data in different places for retransmission or analysis, makes the data susceptible in terms of data theft or manipulation. The government is closely working on these factors and on procedures to guarantee customers’ privacy of information. 

\subsubsection{System resilience against Cyber Attacks}

Cyber security is a very important issue in smart grid due to rising chances of cyber attacks and incidents against this critical sector of the power grid. Cyber security must address deliberate attacks, such as those from disgruntled employees, industrial espionage, and terrorists, as well as inadvertent compromises of the information infrastructure due to user errors, equipment failures, and natural disasters~\cite{Yan13}. From the vulnerability view point, the attacker has the opportunity of entering into a smart grid network, gain access to control software, and change load conditions to destabilize the grid in unpredictable ways. A smart meter is expected to retain its own digital credential and thereby guaranteed to obtain secure connections with the smart meter network. Even if a particular smart meter is attacked, the adversary should be unable to leverage the compromised meter to access information on other meters or penetrate into the AMI of the smart grid.

To address the above cyber security threats, the general requirements for AMI security can be summarized as follows:

\begin{itemize}
\item \textbf{Confidentiality:} Confidentiality as from the perspective of AMI can be perceived as privacy of consumer's consumption pattern and information, i.e., consumption information must remain confidential. This means physical theft of meter to access the stored data, unauthorized access to the data by other connected automated systems through gateways, as well as customer’s access to other customers’ information should be prevented~\cite{Anzalchi2015}. At AMI head end, customer information shall remain confidential and only authorized systems will have access to certain sets of data.
\item \textbf{Integrity:} Integrity in AMI is applicable with respect to the transmitted data from meter to the utility as well as control commands from utility to the meter. Integrity means preventing modifications in the data received from meter, and in the commands sent to the meter~\cite{Yan13}. The objective of the hackers is to breach the integrity of the system, by pretending that they are authorized entities and generate commands for launching attacks. Compared to electromechanical meters, smart meters are resilient against physical or cyber attacks. Smart meters should also be able to detect cyber attacks and ignore all issued control commands to avoid breach in the integrity of the system.
\item \textbf{Availability:} Availability issues in AMI vary according to the type of information communicated in the system. Data that are not critical, can be collected in larger time intervals, and the estimated values be used instead of the actual ones. But, sometimes it is important that the actual values are collected in minimum time. The main reason for unavailability of data in AMI is component failure. Component failure can result from physical damage, software problem, or human tampering with the meter. Communication failure can also be another source of unavailability in AMI. There are several reasons for communication failure such as interference, cut cables, path degeneration, loss of bandwidth, network traffic, etc.
\item \textbf{Accountability:} Means that the entities receiving the data will not deny receiving it and vice versa, i.e., if the entities did not receive the data, they cannot state they have done so. In AMI, accountability is significantly important from financial view point as well as responses to control signals. Accountability requirement in particular is a concern, as different components of an AMI system are generally manufactured by different vendors and owned by different entities. Accurate time stamp of information as well as time synchronization across AMI network is also important in accountability. Audit logs are the most common way to ensure accountability. For accountability with respect to smart meters, all metered values, changes to the parameters and tariffs should be accountable since they are the basis for billing.
\end{itemize}

 \subsubsection{Power Theft Prevention}
Electrical losses can occur at any stage of generation, passing
through step-up transformers, transmission, distribution, and utilization. Generally, losses during generation are technically easy to define, while losses in transmission and distribution are difficult to quantify. Another category of losses can be in the form of technical loss
and non-technical loss. Technical loss is considered natural because of power dissipation in lines and components. On the other hand, non-technical loss during transmission and distribution of electricity is difficult to detect, calculate and prevent, thus, resulting in a major problem for utility. The electro-mechanical meters used in traditional systems for metering purposes provide very less or no security and are easy to manipulate. Thefts in electro-mechanical meters are identified using methods~\cite{Anas2012} such as direct connection to distribution lines, grounding the neutral wire, attaching a magnet to electromechanical meter 
etc. Smart meters usage resulted in elimination or reduction in the above mentioned issues of electro-mechanical meters. Smart meters are capable of recording zero readings and informing the utility companies through AMI.

There also exist complex techniques for power theft that do not involve the meter directly. Current Transformer tampering is one of them. Current transformers are responsible for matching the grid current rating with the meter rating for meters of large loads. The secondary side wires of current transformers are generally insulated, but it is possible to damage this insulation and tap these wires. Based on the number of wires tampered, the meter is compelled to read less or even zero current amounts. The other method is to exchange the position of damaged wires, causing phase shift and altering the meter reading. Some of the stealing techniques used in electro-mechanical meters work in systems with smart meters and AMI too. Data tampering can occur at three different stages: during data collection, when data is stored in the meter and when data transits across the network. Data tampering during collection can happen with both conventional and smart meters. Interfering with data at the other two stages can only happen with smart meters. In comparison with conventional systems, AMI makes tampering meters more difficult using data loggers. The loggers are capable of recording power outages to the meter or inversion of power flow. Attackers planning to use inversion or disconnecting techniques need to also erase the logged events stored in the meter. Therefore, its removal falls into the second category of tampering with stored data in the meter. If attackers access the stored data of smart meter they will have complete control over the meter as the time of use tariffs, received or executed commands, event logs,
consumption and time stamps and the firmware reside there. Generally, in power theft, the firmware and whole stored data in the meter is not the attackers interest, instead manipulating stored total demand and auditing logs is sufficient for them. In another scenario, the data is modified while it is being transferred over the network. This comprises of injecting false data into the system, or intercepting communications within the infrastructure. This type of attack is possible at each node of the infrastructure. Different techniques were developed and introduced to estimate and locate power theft. 
 
\subsection{Role of Key Management Systems in AMI}
As mentioned in Section I, AMI is a new emerging technology for smart grid, and is defined as the system used to measure, collect, store, analyze, and use energy usage data~\cite{Xia12}. It also facilitates in building a bridge between consumers and electric power utilities. For providing the widest possible platform for delivering a wide range of applications in the future, open network and information techniques introduced to the smart grid increases the chances of cyber security threats. Generally, the cyber security requirements of AMI include confidentiality, integrity, and availability. Before AMI can be deployed, the confidentiality for customer privacy and customer behavior, as well as message authentication for meter reading, DR and load control messages, are the major security requirements that need to be provided. Confidentiality and integrity are solved by encryption and authentication protocols, that depend on the security of cryptographic keys. To ensure the security, the key management for large amounts of devices in AMI systems is very important. The key management system always comprises of a key organizational framework, key generating, refreshing, distribution, storage policies, etc~\cite{Liu13}.
\subsubsection{Key Management in Different Transmission Modes}
Messages in AMI can be classified into three classes based on their mode of transmission: unicast, broadcast and multicast.
\begin{itemize}
\item \textbf{Unicast communication:} used for messages that are transmitted from one point to the other one, for example, when a smart meter reports its power consumption statistics and estimated future energy demand to the utility system. Based on the different messages in AMI systems, the unicast transmission mode consists of three types of messages: meter data, joining or leaving of Demand Response projects and remote load control. The messages are bidirectional: from user level side to the management side or vice-versa.
\item \textbf{Broadcast communication:} is used when messages are transmitted from one point to all the other points. A typical example of a broadcast notification message includes the real-time electricity pricing information sent from utility system to all smart meters. Two types of messages are transmitted in broadcast mode, namely, publishing of DR projects and electrical pricing information. From the key management outlook, the session keys should be refreshed before every broadcast session to provide confidentiality and integrity of the messages. The key refreshing policy for broadcast communication necessitates that the key should be refreshed periodically.  
\item \textbf{Multicast communication:} is used when messages are transmitted from one point to a subset of the other points, e.g., a remote load control message from utility system to SMs that subscribes to the same DR project. For all the types of AMI messages, electrical pricing information and remote load control are transmitted using multicast mode. As the users who subscribe to a DR project are not fixed, the group members who receive the multicast messages should be updated in a certain period (such as one day and one week) depending on the actual situation of electric power utilities. Therefore, the key management for multicast communication is separated into two parts. One is similar to the broadcast communication. The session key of multicast communication also should be generated before every new session. In another one, considering that the users may join or quit a DR project, the group key and additional values should be regenerated and refreshed with the support of unicast communication. 
\end{itemize}

\section{Key Management Approaches}
\begin{figure*}
 \centering 
 \includegraphics[width=6in]{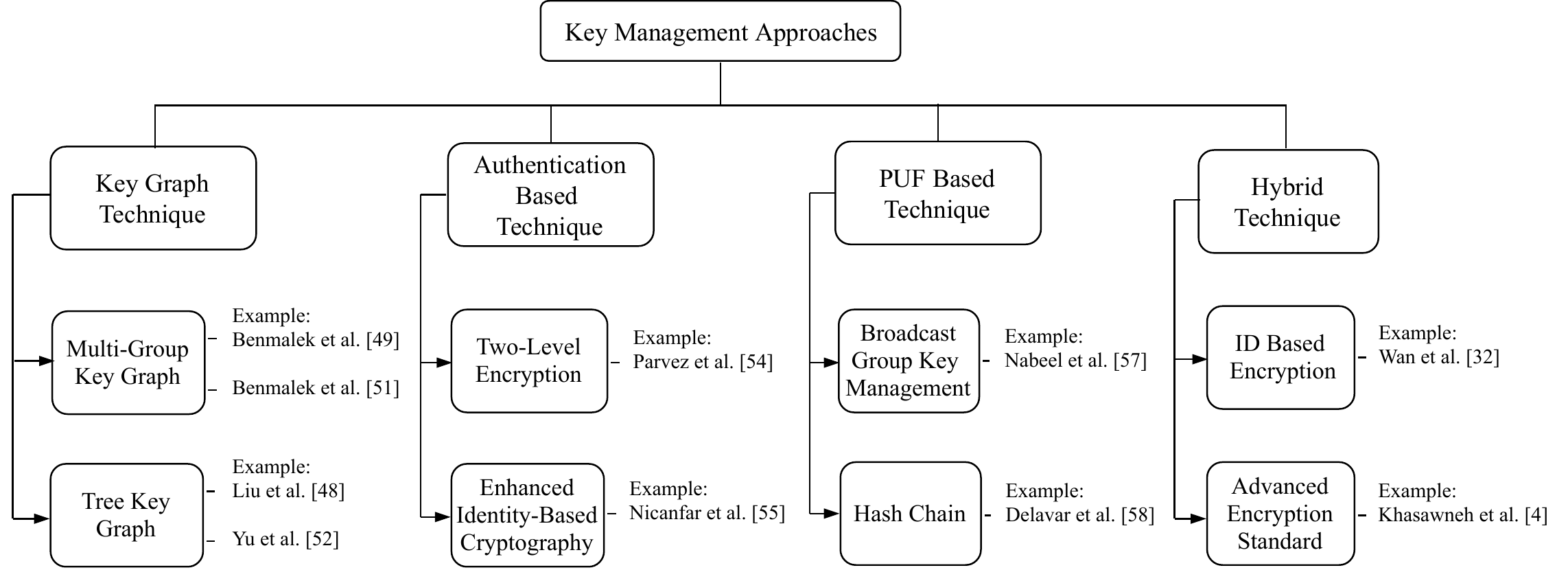}
 \caption{Taxonomy for Key Management Approaches in AMI.}
 \label{fig:sample1}
\end{figure*}

According to \cite{Wan14}, key management systems are an important part of AMI that facilitates secure key generation, distribution and rekeying. Different approaches were adapted for ensuing efficient key management. In the literature authors reported works that deal with the issue of key management system. All these works are conducted through different approaches based on different secure key generation and distribution mechanisms. In this survey we categorize the existing works, mainly into four categories, namely, key graph technique, encryption based technique, Physically Unclonable Function (PUF) based technique and hybrid technique. In Fig.~\ref{fig:sample1}, we provide a classification of the state-of-the-art works in the area of KMS in AMI of smart grids, which we have analyzed in the following sections.   

\subsection{Key Graph Technique}

By far, key graph technique is the most commonly used key management mechanism due to the ease of implementation and efficient performance. The key graph technique can be classified into two categories, namely, multi-group key graph technique and tree key graph technique. The following subsections provide description of the works under the above mentioned classification. 

\subsubsection{Multi-Group Key Graph}
This section illustrates the state-of-the-art works that adapted multi-group key graph technique for secure key management in AMI. 

\par Authors in~\cite{Benmalek2015} proposed a key management scheme for AMI based on an efficient and scalable multi-group key graph technique for securing unicast, multicast, and broadcast communications. This work also safeguards the security requirements of AMI. The scheme is based on a multi-group key graph structure that supports the management of multiple demand response projects concurrently for every customer. Here, individual keys between the smart meters and MDMS is established using a specific method of securely exchanging cryptographic keys over a public channel. The individual keys are refreshed periodically and used in two ways. The first one secures unicast communication between MDMS and SMs, while the second is used for generating the multi-group key graph for secure multicast communications. The MDMS is responsible for generating a group key that is also refreshed periodically for the DR project. The group key is generated and transmitted through secure channels for each SM and used for securing broadcast messages. 
\par The scalability issue is addressed using Logical Key Hierarchy (LKH), where a key tree is used for every DR project. In LKH, each member keeps a copy of its leaf secret keys and all other keys corresponding to the nodes in the path from its leaf to the root. The authors demonstrate that using their proposed LKH, scalability is ensured for large smart grids with dynamic demand response projects. Also, to reduce storage and communication costs in key management, a multi-group key graph structure is proposed in this work. The proposed key graph technique allows multiple DR projects to share a new set of keys. When a user joins or leaves a DR project, the communication cost for rekeying operations does not increase significantly compared to the cost induced by using separate LKH tree inside each DR project. In this work the multi-group key graph structure is modeled as a two level graph. In the lower level, every LKH tree denotes a set of users with the same first DR project subscription, where the leaf node of the tree is a user's individual key and the tree's root is the DR project's group key. The graph in the upper level represents combinations of root keys for users subscribing to multiple DR projects at the same time. The multi-group key graph used in this work has the following properties: (i) a user only belongs to one LKH tree in the multi-group key graph corresponding his/her first DR project subscription. The user also has a copy of its leaf secret key and all keys corresponding to the nodes in the path from its leaf to the root in this tree; (ii) a user has all group keys of the other DR projects to which he/she is subscribed; and (iii) if a user leaves his/her first DR project and remains subscribed to one or more DR projects, he/she will shift to a new LKH tree. Authors claim that all these features ensure a user does not subscribe and pay for the same DR project multiple times.

In~\cite{Benmalek15}, the authors proposed a scalable multi-group key management for AMI for securing data communications in AMI. It is a key management scheme that supports unicast, multicast and broadcast communications based on an efficient multi-group key graph technique. The multi-group key graph structure supports the management of multiple DR projects simultaneously for each customer. Also, authors demonstrate that this new structure scales to large smart grids with dynamic DR projects membership while meeting smart meters’ constraints in terms of memory and bandwidth capacities. 
\par In this work, a particular method of securely exchanging cryptographic keys over a public channel is used for establishing individual keys between the MDMS and smart meters. The individual keys are refreshed periodically and utilized in two ways. The first one is used for securing unicast communications between MDMS and the SMs, and the other one is used for generating the multigroup key graph for secure multicast communications. The MDMS must also generate a group key that is refreshed periodically for the default DR project. This group key is generated and transmitted through secure channels for each SM, and used for securing messages transmitted in broadcast mode. 
\par To address the scalability issue, the key graph techniques used in this work is One-Way Function Trees (OFT) that is improvised over the LKH protocol. In OFT, the MDMS and all users individually compute the group key. The keys of interior nodes are recursively computed from the keys of their children. For reducing storage and communication costs in key management, a multi-group key graph structure is proposed in this work. The key graph technique allows several DR projects to share a new set of keys. The multi-group key graph structure is modeled as having two levels, lower and upper levels. In the lower level, each OFT represents a set of users with the same first DR project subscription. The leaf node of the tree represents a user’s individual key and tree’s root is the DR project’s group key. The graph in the upper level represents combinations of root keys for users subscribing to multiple DR projects at the same time. The proposed key management scheme supports both backward secrecy and forward secrecy. An analysis of security and performance, and comparison results exhibit that the scheme induces low storage as well as low communication overheads.
\par Authors in~\cite{Ben18} proposed four key management schemes that can simultaneously support security, scalability, efficiency and versatility. The first scheme named as Versatile and Scalable key management scheme for AMI (VerSAMI), ensures secure unicast, multicast and broadcast communications for a large-scale AMI system. VerSAMI also supports the management of multiple DR programs. This is done so that many customers are able to subscribe to multiple DR programs simultaneously and subscribe/unsubscribe to any DR program at any time. The multi-group key graph technique used here efficiently handles the rekeying operations, while meeting the limitations of smart meters in terms of memory and bandwidth capacities. An improved version of VerSAMI, called, VerSAMI+ is also proposed by the authors, that provides enhancement in communication overhead. The problems in VerSAMI as well as VerSAMI+ that occured due to individual rekeying, and also to reduce the number of rekeying operations, another variant of VerSAMI, called Batch-VerSAMI was proposed by the authors. In Batch-VerSAMI, membership changes are handled in batches instead of handling individually. The authors additionally proposed a dynamic membership model that simulates the AMI system behavior. To prove the efficiency of the proposed schemes, security and performance analyses, as well as simulations are performed with state-of-the-art schemes.

\subsubsection{Tree Key Graph}
This section describes the schemes that use the tree key graph as the structure for the key management system in AMI.
\par In~\cite{Liu13}, authors design a key management system for dealing with the security requirements in AMI. The authors develop the key management framework of the AMI system based on the key graph. Three different key management processes are designed for supporting the hybrid transmission modes that also include key management for unicast, broadcast, and multicast modes. For minimizing the storage and computation constraints of SMs, simple cryptographic algorithms are chosen for key generation and refreshing policies. As the members in a particular DR project is not consistent, specific key refreshing policies are designed. Mainly the key management framework based on the key graph concentrates on managing the keys of a large number of SMs. Finally, the security and performance of the KMS are analyzed. According to the results, the proposed scheme is a possible solution for AMI systems. Though the functions of AMI are not fixed, but a KMS is designed for the collection of possible functions. In actual scenario, the users can choose part of the KMS for particular applications. Also, the communications are not fixed but the characteristics of messages transmitted in the communication channels is decided by the function requirements. The authors design three types of key management processes for unicast, broadcast, and multicast communications depending on function requirements and message types. Specific key regeneration and refreshing policies are designed in each process considering time requirements of functions, computation, and storage limitations of SMs.

\par In~\cite{Yu15}, the authors introduce Information Centric Networking (ICN) in AMI systems and also proposed a key management scheme for large number of smart meters for ensuring confidentiality, integrity and authentication. The scheme is designed with an objective to control network congestion, support mobility and ensure security. Authors claim that ICN's self-contained data security can help AMI system simplify security processing. The energy data in AMI system needs to be kept secret as they can reflect the privacy of people, such as, someone is at home or not and eating habits. The ICN implementations ensure data integrity as well as authenticity by signature and confidentiality by encryption at data creation, that differs from the protection available in end-to-end communication channels. It relies on the protection of data itself. Here, secure message exchange for unicast, broadcast and multicast is ensured. 
\par The ICN-AMI system structure develops the KMS frame structure by key graph similar to that of \cite{Wong00}. The user key is used for the unicast mode, the group key is used for the multicast mode and the root key is used for the broadcast mode. The ICN's self-contained digital signature provides integrity and authenticity and the proposed KMS provides data confidentiality by encrypting payload for the three different transmission modes. For the unicast transmission mode, four types of messages are used namely, remote load control, join or leave DR projects, loss or restoration of power notification and the meter reading data. There are three steps in ICN-KMS for unicast transmission mode to ensure data confidentiality, integrity, availability and authenticity. In the ICN process, the self-contained data security provides integrity, authenticity, and availability. Messages having electrical pricing information and released DR projects use broadcast mode. Both these two types of messages are sent from operation domain to customer domain. In broadcast transmission mode, confidentiality is provided by root session key and the ICN proves the integrity, availability, and authentication by signature. The multicast transmission mode includes two types of messages: remote load control and electrical pricing information. The messages are transmitted from operation domain to customer domain. As the users participating in DR projects can change at any time, anybody can choose to join or leave the project at any time based on their personal preference. The security and performance analysis validate the correctness of the authors claims. The simulation results also demonstrate that ICN-AMI system achieves better performance with respect to resisting attacks, KMS design, congestion control and mobility support.

\subsection{Authentication Based Technique}
The key management system in AMI using encryption based technique generally follows two level encryption and identity based cryptography mechanisms. The following sections discusses works that used authentication based technique for KMS in advanced metering infrastructure.

\subsubsection{Two level encryption}
This section describes the work based on two level encryption for key management system in AMI.  
\par In~\cite{Parvez17}, a key management scheme is proposed based on two level encryption method. The encryption is based on two partially trusted simple servers that implement this method without increasing packet overhead. One server is responsible for data encryption between the meter and control center and the other server manages the random sequence of data transmission. Authors further introduce one-class support vector machine algorithm for node-to-node authentication utilizing the location information and the data transmission history (node identity, packet size and frequency of transmission). This mechanism helps in securing data communication privacy without increasing the complexity of the conventional key management scheme. This work is an extension of one of the earlier work's of the authors. The extended work uses Received Signal Strength (RSS) and One Class Support Vector Machine (OCSVM) techniques for node to node authentication. RSS is used for localization of meters using the received signal strength from neighbouring meters. OCSVM is used for detecting new and outlier data/packet, using current and previous data transmission history. Further, OCSVM based node authentication increases robustness in key management system without increasing any overhead and is easy to implement in devices with limited memory and computational ability. The introduction of two separate servers for key management and random sequenced packet transmission enhances robustness in security in untrustworthy communication medium and servers. Both qualitative and quantitative analyses reveal the scheme's efficacy in providing improvements in the key management schemes of AMI.

\subsubsection{Identity based cryptography}
This section illustrates the work that use identity based cryptography as the basis for KMS.
\par This paper~\cite{Nicanfar14} proposes an efficient scheme that mutually authenticates a smart meter of a home area network and an authentication server in smart grid. This work also proposes a key management protocol based on the Enhanced Identity-Based Cryptography (EIBC)~\cite{Nicanfar12} for secure smart grid communications using the public key infrastructure. EIBC is a modified version of the IBC (Identity-Based Cryptography) scheme, where the enhanced version significantly reduces the overhead of key renewals. The key management protocol employs PKI for smart grid communications as specified by NIST. Also, the network overhead caused by control packets for key management is minimized. The efficiency results of the key refreshment protocol demonstrates that the Security and Authentication Server (SAS), periodically broadcasts a new key generation to refresh the public/private key pairs of all the nodes as well as any required multicast security keys. The authors claim that their mechanisms are capable of preventing several attacks such as brute force, replay, DoS etc. while reducing the management overhead. The improved efficiency for key management is realized by periodically refreshing all public/private key pairs as well as any multicast keys in all the nodes using only one newly generated function broadcast by the key generator entity. Security and performance analyses are presented that demonstrate the desirable attributes of the presented scheme.

\subsection{PUF Based Technique}
In contrast to key graph and hybrid techniques mentioned above, PUF based technique for secure key management in AMI has not been exploited much in the literature. This section mentions two such works in the area of PUF based KMS in AMI. One of them is based on broadcast group key management~\cite{Nabeel15} while the other uses hash chain~\cite{Delavar17}.

\subsubsection{Broadcast Group Key Management} This section provides description for the work that utilizes the broadcast group key management technique in PUF base KMS. 
\par In~\cite{Nabeel15}, the authors provide an end to end security for AMI networks. The model is based on using weak (PUFs)~\cite{Nabeel2012} as a security primitive to provide strong hardware based authentication for smart meters and collectors. During the initialization phase, the utility generates three hash functions and a pair of challenges. The challenges are used for triggering PUF responses, whereas, the hash functions are used for obtaining hash codes and computing smart meter symmetric key. The proposed approach provides efficient solution to manage keys and a robust authentication mechanism. The solution is developed using PUF devices that are cheap to manufacture and provide hardware based strong authentication mechanisms against spoofing attacks. The access level passwords for smart meters are generated and regenerated by use of the PUF devices' hardware based one way function. Strong defense against key leakage is provided by the PUF based secret generation method as the key is not stored in memory. For supporting multicast communications, this work utilizes Broadcast Group Key Management (BGKM) scheme~\cite{Zou08},~\cite{Shang10}. The BGKM scheme is a special type of group key management scheme that allows a subset of nodes to communicate efficiently. BGKM is not computationally intensive as only symmetric key cryptography is used. Every smart meter in the network is designated an unique secret that allows the smart meter to derive the group key. The BGKM scheme is efficient and scalable as it is capable of addressing any subset of nodes with a single message.

\subsubsection{Hash Chain} The work described in this section employs hash chain mechanism for PUF based KMS.
Authors in~\cite{Delavar17} devised mechanisms for authenticated key exchange protocol and message broadcasting protocol utilizing the PUF technology in communication parties. The PUF is embedded in the head end system for key exchange protocol while in message broadcasting protocol, it is embedded in the smart meters. The PUFs are used for generating the commitments and random numbers that are required for running the protocols. In the authenticated key exchange protocol, the head end system and smart meter exchange a session key after authenticating each other. The protocol consists of four phases namely, initialization, registration, mutual authentication and key exchange. We utilize hash chain idea and a modified version of Schnorr protocol~\cite{Seurin2012} for proposing an authenticated broadcast messaging protocol for the AMI systems. In the proposed protocol, the head end is authenticated by the smart meters. Thereby, the smart meters are ensured that the message received is broadcast by the head end. The message broadcasting protocol consists of three phases, initialization, registration and authentication. The security analysis of the proposed authenticated key exchange protocol shows that it provides all the requirements of a secure one. Also, we showed that our authenticated message broadcasting protocol is secure against corrupted Smart Meters.     

\subsection{Hybrid Technique}
Hybrid techniques for KMS provide for both symmetric and asymmetric encryption for securing AMI in smart grids. Hybrid techniques are grouped under two categories, namely, ID based encryption and advanced encryption standard. This section provides description on potential works that deal with KMS using the hybrid techniques.

\subsubsection{ID based encryption} The work in the following section describes how ID based encryption assist in hybrid key management in AMI.
\par The work in~\cite{Wan14} proposes a hybrid key management scheme for AMI by combining symmetric cryptosystem and public key cryptosystem. The scheme employs elliptic curve cryptosystems for achieving efficient authentication and session key generation. Also, the scheme makes use of a specially designed key hierarchy for efficient generation and updation of group keys. The key management scheme is based on ID-based encryption and also employs the key graph technique for efficient multicast key management. The main benefit of using ID-based encryption is that no public key certificate is required and the public key is simply the identity of an entity. This work is mainly involved with two security issues: end-to-end key establishment and multicast key management. End-to-end key establishment enables the management side, i.e., MDMS to establish session keys with each smart meter securely, and the resulting session keys are used in multicast key management for scalable and efficient group key management for AMI. Security analysis and performance evaluation for this scheme reveal that it is secure and efficient for AMI in smart grid.
\begin{figure}[b]
 \centering 
 \includegraphics[height=61mm, width=70mm]{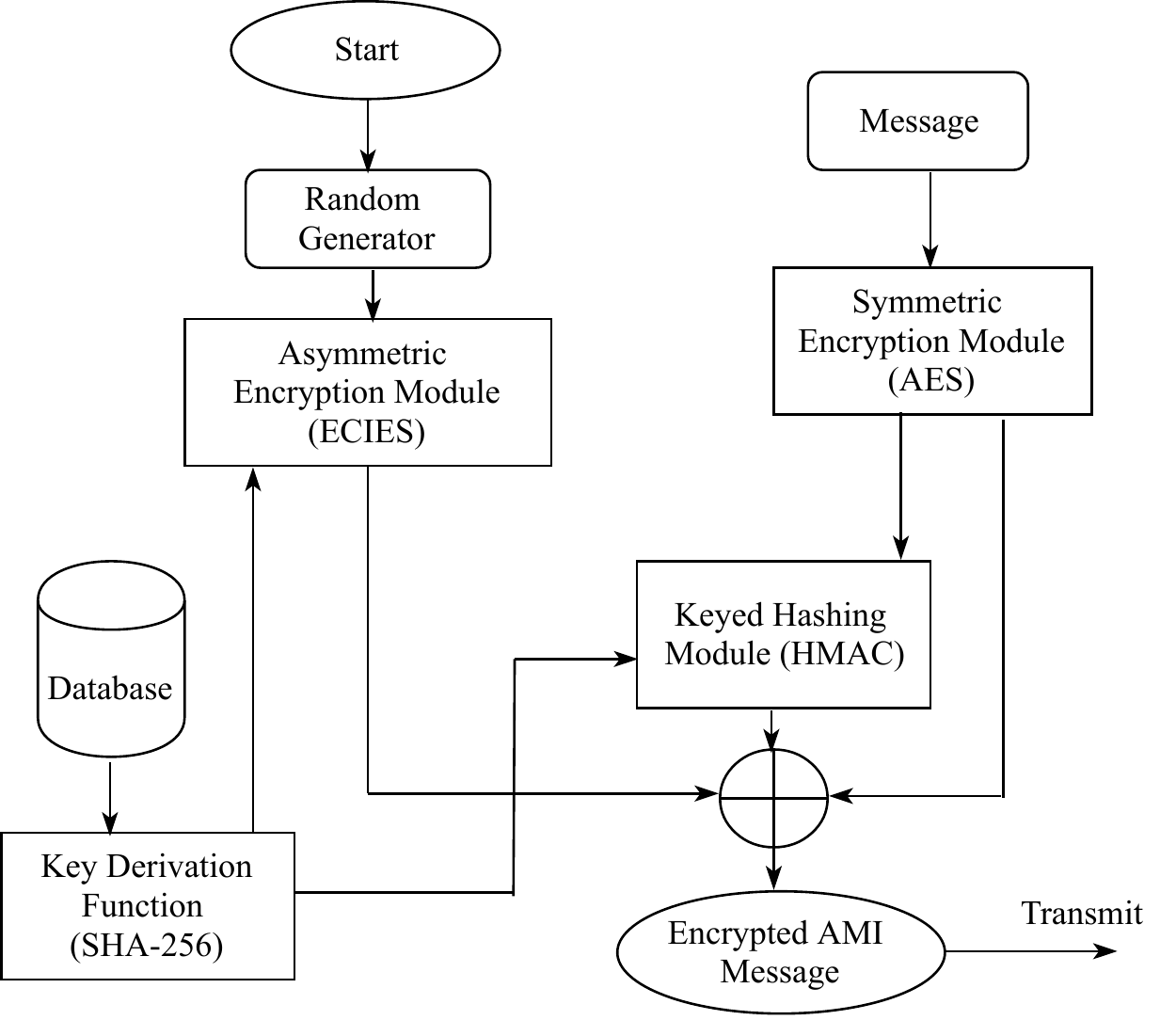}
 \caption{Hybrid Cryptosystem.}
 \label{fig:Fig_4}
\end{figure}

\subsubsection{Advance encryption standard}
The work in this section employs advance encryption standard for key management systems in AMI.
\par Authors in~\cite{Khasawneh17} proposed a hybrid encryption scheme as shown in Fig.~\ref{fig:Fig_4}, that exploits symmetric and asymmetric encryptions for securing smart metering network. The proposed method employs elliptic curve cryptography, where a precomputation stage is introduced to minimize the computation overhead of elliptic curve encryption by eliminating the time required to perform the scalar-point multiplication. In this work a hybrid encryption scheme is developed that exploits the secret key encryption with the public key cryptography. The proposed scheme provides a lightweight key management scheme to reduce the overhead of key generation, distribution and renewal to cope with AMI constraints. The scheme uses Advance Encryption Standard (AES) for the data encapsulation cryptosystem and Elliptic curve encryption for the key encapsulation cryptosystem. The proposed hybrid encryption cryptosystem is composed of three units: symmetric encryption module, asymmetric encryption module and message integrity module. The symmetric encryption module is used to encrypt AMI messages using AES-128 with an arbitrary key produced by random generation unit. The asymmetric encryption module is used to encrypt the arbitrary key used by the symmetric encryption module. The symmetric encryption module acts as data encapsulation system, while the asymmetric module acts as key encapsulation system. Message integrity module is used to generate the integrity code that enables detection of any form of tampering with the secure message.

\par The messages transmitted using this proposed protocol has the packet format as follow: a payload field that stores the information that needs to be sent to the destination. The synch and clock tolerance fields are used to protect against reply attack. The synch field consists of 128 bits that stores the message timestamp corresponding to message creation time, whereas clock tolerance (32 bits) indicates for how long (in milliseconds) the message remains valid. When a message is received, the synch field is compared with the current time. If the difference between the current time and synch is greater than the value of the clock tolerance, the message is considered replayed and is ignored. The precomputation parameter is used to improve the security of the proposed protocol by frequently updating the precomputed values that are used to derive the cryptographic keys. The input to key derivation function is updated once a new precomputation parameter is received by multiplying the received value by the private key. The hybrid cryptosystems encrypt the arbitrary key and appends it to the message. The encrypted key is stored in the key field. The size of this field depends on the original key size and the public key algorithm used to encrypt the key. Message Integrity Code (MIC) is used to protect message integrity and authenticity. A message that is tamped with or originated from an unauthorized source is detected when MIC field is checked. The results obtained demonstrate that the scheme performs well in terms of computation overhead, storage requirements and communication overhead. Also, the scheme ensures message confidentiality, thereby, providing defense against eavesdropping and traffic analysis attack. 

\begin{table}[b]
\centering
\caption{Notation Table}
\label{table 1}
\scalebox{0.75}{
\begin{tabular}{|l|l|}
\hline
\textbf{Notation} & \textbf{Description} \\ \hline
$n$        & Number of smart meters  \\ \hline
$N_{pr}$        & Number of DR projects           \\ \hline
$m_j$        & Number of jth DR project member           \\ \hline
$Nsub(u_i)$        & The number of DR projects to which subscribes user $u_i$           \\ \hline
$|K|$        & Size of the key in bits           \\ \hline
$X$        & $Nsub(u_i)$           \\ \hline
$Y$        & $N_{pr}-Nsub(u_i)$           \\ \hline
$c$        & $\log_2(N_{pr})$          \\ \hline
$h_k$        & Height of the new home DR project          \\ \hline
$C_P$        & Cost for bilinear pairing           \\ \hline
$C_M$        & Cost for the multipoint multiplication           \\ \hline
$C_\varepsilon$        & Cost for the encryption function $\varepsilon$           \\ \hline
$C_r$        & Cost for generating one key            \\ \hline
$C_f$        & Cost for evaluation of the one-way function           \\ \hline
$A$        & $C_P+C_M$           \\ \hline
$B$        & $C_\varepsilon+C_f$  \\ \hline
$P$        & $(4n+5)+C_f$           \\ \hline
$Q$        & $4nC_f$            \\ \hline
\end{tabular}
}
\end{table}

\begin{table*}[h]
\caption{Security Analysis}
\label{table 2}
\begin{center}
\centering
\scalebox{0.75}{
\begin{tabular}{|c|c|c|c|c|c|c|c|}
\hline
\textbf{Schemes} & \textbf{Key Generation} & \textbf{Key Sharing} & \textbf{Key Freshness} & \textbf{Forward and Backward Security} & \textbf{Confidentiality} & \textbf{Authentication} & \textbf{Integrity} \\
\hline
\textbf{MK-AMI} \cite{Benmalek2016} & \checkmark & \ding{53} & \checkmark & \checkmark & \ding{53} & \ding{53} & \ding{53} \\
\hline
\textbf{eSKAMI} \cite{Benmalek15} & \checkmark & \ding{53} & \checkmark & \checkmark & \checkmark & \ding{53} & \checkmark\\
\hline
\textbf{KMSSC} \cite{Liu13}  & \checkmark & \checkmark & \checkmark & \checkmark & \checkmark & \checkmark & \checkmark\\
\hline
\textbf{KMSSC-IC} \cite{Yu15} & \checkmark & \ding{53} & \ding{53} & \checkmark & \checkmark & \checkmark & \checkmark\\
\hline
\textbf{SEES} \cite{Nabeel15} & \checkmark & \ding{53} & \checkmark & \checkmark & \checkmark & \checkmark & \checkmark\\
\hline
\textbf{SKM} \cite{Wan14} & \checkmark & \ding{53} & \checkmark & \checkmark & \checkmark & \checkmark & \checkmark\\ 
\hline
\textbf{SAMI} \cite{Benmalek2015} & \checkmark & \checkmark & \checkmark & \checkmark & \checkmark & \checkmark & \checkmark\\
\hline
\end{tabular}
}
\end{center}
\end{table*}

\begin{table*}[htbp]
\begin{center}
\centering
\caption{Computation Cost}
\label{table 4}
\scalebox{0.92}{
\begin{tabular}{|c|c|c|c|c|c|c|c|c|}
\hline
\multirow{2}{*}{\textbf{Schemes}} & \multicolumn{2}{c|}{\textbf{\begin{tabular}[c]{@{}c@{}}End to End Key\\ Establishment\end{tabular}}} & \multicolumn{2}{c|}{\textbf{\begin{tabular}[c]{@{}c@{}}Initializing a Group\end{tabular}}} & \multicolumn{2}{c|}{\textbf{\begin{tabular}[c]{@{}c@{}}Adding a Member\end{tabular}}} & \multicolumn{2}{c|}{\textbf{\begin{tabular}[c]{@{}c@{}}Deleting a Member\end{tabular}}} \\ \cline{2-9} 
                                  & \textbf{MDMS}                             & \multicolumn{1}{l|}{\textbf{SM}}                            & \textbf{MDMS}                                  & \textbf{SM}                                  & \textbf{MDMS}                                & \textbf{SM}                               & \textbf{MDMS}                                & \textbf{SM}                                \\ \hline
\textbf{SAMI} \cite{Benmalek2015}                     & $n(A)$                                         & $A$                                                           & $n(A)+nC_r$                                              & $hC_\varepsilon$                                            & $h(B)+C_r$                                            &  $h(B)$                                           & $h(B)+C_r$                                            &  $h(B)$                                        \\ \hline
\textbf{KMSCC} \cite{Liu13}                    & -                                         & -                                                           & $nC_\varepsilon$                                              & $C_\varepsilon$                                            & $P+(n+2)C\varepsilon$                                            & $Q+C_\varepsilon$                                         & $P+(n+2)C\varepsilon$                                             & $Q+C_\varepsilon$                                          \\ \hline
\textbf{SKM} \cite{Wan14}                      & $n(A)$                                         & $A$                                                           & $2n(B)+nC_r$                                                & $hC_\varepsilon$                                             & $h(C_\varepsilon+2C_f)+C_r$                                             & $h(B)$                                         & $h(C_\varepsilon+2C_f)+C_r$                                             & $h(B)$                                          \\ \hline
\textbf{eSKAMI} \cite{Benmalek15}                   & $n(A)$                                           & $A$                                                           & $n(B)+nC_r$                                              & $hC_\varepsilon$                                            &  $h(B)+C_r$                                               &  $h(B)$                                            &  $h(B)+C_r$                                               &  $h(B)$                                             \\ \hline
\textbf{MK-AMI} \cite{Benmalek2016}                   & $n(A)$                                           & $C_P+C_M$                                                            & $n(B)+nC_r$                                              & $hC_\varepsilon$                                             & $h(B)+C_r$                                             & $h(B)$                                          & $h(B)+C_r$                                             & $h(B)$                                          \\ \hline
\end{tabular}
}
\end{center}
\end{table*}

\begin{table*}
\begin{center}
\centering
\caption{Communication Cost}
\label{table 5}
\begin{tabular}{|c|c|c|c|c|}
\hline
\multirow{3}{*}{\textbf{Schemes}} & \multicolumn{4}{c|}{\textbf{Communication Overhead}}                                          \\ \cline{2-5} 
                                  & \multicolumn{2}{c|}{\textbf{Member Addition}} & \multicolumn{2}{c|}{\textbf{Member Deletion}} \\ \cline{2-5} 
                                  & \textbf{Unicast}     & \textbf{Broadcast}     & \textbf{Unicast}     & \textbf{Broadcast}     \\ \hline
\textbf{SAMI} \cite{Benmalek2015}                     & $h_j|K|+c$                      & $|K|+c$                      & ($h_j+N_{pr}-1)|K|$                    & ($h_j+2X+Y+h_k)|K|+c$                      \\ \hline
\textbf{KMSCC} \cite{Liu13}                     & $2n|K|$                    & $0$                      & $2n|K|$                    & $0$                      \\ \hline
\textbf{SKM} \cite{Wan14}                      & $h|K|$                    & $h|K|+h$                      & $0$                    & $h|K|+h$                      \\ \hline
\textbf{eSKAMI} \cite{Benmalek15}                   & $2h_j|K|+c$                    & $|K|+c$                       & $(2h_j+N_{pr}-1)|K|$                    & $(2h_j+2X+Y+2h_k)|K|+c$                      \\ \hline
\textbf{MK-AMI} \cite{Benmalek2016}                   & $h_j|K|+c$                    & $|K|+c$                      &  $(h_j+N_{pr}-1)|K|$                   & $(h_j+2X+Y+h_k)|K|+c$                      \\ \hline
\end{tabular}
\end{center}
\end{table*} 

\section{Comparative Study} 
The objective of this section is to provide a detailed analysis of the works that carried out efficient key management system of AMI in smart grids, in the perspective of both security and performance analyses. In Table~\ref{table 1}, we summarize the terminologies that are used throughout the remaining of this paper.

\par Table~\ref{table 2} highlights the security analysis of the various schemes. In this table, we provide an insight into how efficient the schemes are in comparison to each other by taking into consideration certain security parameters. The parameters considered are key generation, key sharing, key freshness, forward and backward security, confidentiality, authentication and integrity. On the other hand, performance analysis of the different schemes through computation cost, communication cost and storage cost are illustrated using three tables. 

In Table~\ref{table 2}, the processes of key generation, key sharing and key refreshment reflect that low communication overhead is induced, that is important for time critical scenarios in AMI. Key refreshing depends on whether there are users joining or quitting the DR project. If there is an user who wants to join or quit the DR project, the group key should be refreshed at the update time. The forward secrecy implies that previously used secret keys and messages must be inaccessible by the new users who participate in a DR project, while the backward secrecy means that the future secret keys and messages must be inaccessible by users who leave a DR project. Forward secrecy (in the context of group key management) means that evicted members cannot learn the new group key. Backward secrecy (in the context of group key management) means that newly added members cannot learn previous group keys. As users participating in DR projects are not fixed and can join or leave any DR project at any time, therefore, preserving forward and backward secrecy should be guaranteed. 

Table~\ref{table 4} and Table~\ref{table 5} provide the comparative analysis of computation and communication costs, respectively of the various schemes. 
The computation cost of the key management protocols are divided into four parts: end to end key establishment, initializing group, adding a member and deleting a member for both MDMS and smart meter. Considering one example from Table~\ref{table 4}, the computation cost of the scheme SKM, for initializing a group, is higher than that of the scheme KMSSC. The results of Table~\ref{table 4} also reflect that, adding a member or evicting a member for the SKM scheme is much more efficient in comparison to KMSSC scheme. The cost for the end-to-end key establishment protocol for the scheme KMSSC is unavailable, as it is not specified what key establishment mechanism is used. 
\par The benchmarks provided in Table~\ref{table 5}, demonstrate the communication overhead of the different schemes. The comparative study is done using unicast and broadcast communication for both member addition as well as member deletion. Taking one example from Table~\ref{table 5}, the results reveal that the scheme MK-AMI performs better than the scheme eSKAMI in terms of communication cost.     
\par Table~\ref{table 3} provides the storage cost for the different schemes. The storage costs are calculated for both MDMS and smart meters. While the storage cost for MDMS is the same for all the schemes except KMSCC, the same is also true for the storage overhead in smart meters, except the schemes, KMSCC and SKM.

\begin{table}[h]
\centering.
\caption{Storage Cost}
\label{table 3}
\scalebox{0.80}{
\begin{tabular}{|c|c|c|}
\hline
\multirow{2}{*}{\textbf{Schemes}} & \multicolumn{2}{c|}{\textbf{Storage Overhead}} \\ \cline{2-3} 
                                  & \textbf{MDMS}          & \textbf{SM$_i$}          \\ \hline
\textbf{SAMI} \cite{Benmalek2015}                     & $2\sum\nolimits_{j=1}^{N_{pr}} \left(m_j-1 \right) +1$                         & $\log_2\left(|A|\right)+N sub(u_i)+1$                      \\ \hline
\textbf{KMSCC} \cite{Liu13}                   & $n+N_{pr}+1$                      & $Nsub(u_i)+2$                      \\ \hline
\textbf{SKM} \cite{Wan14}                      & $2\sum\nolimits_{j=1}^{N_{pr}} \left(m_j-1 \right) +1$                         & $\sum\nolimits_{j=1}^{Nsub({u_i})} \left(\log_2m_j+1 \right) +1$                    \\ \hline
\textbf{eSKAMI} \cite{Benmalek15}                   & $2\sum\nolimits_{j=1}^{N_{pr}} \left(m_j-1 \right) +1$                         & $\log_2\left(|A|\right)+N sub(u_i)+1$                     \\ \hline
\textbf{MK-AMI} \cite{Benmalek2016}                   & $2\sum\nolimits_{j=1}^{N_{pr}} \left(m_j-1 \right) +1$                         & $\log_2\left(|A|\right)+N sub(u_i)+1$                     \\ \hline
\end{tabular}}
\end{table}

\section{FUTURE DIRECTIONS}
This section presents some areas related to smart grids in particular with key management in AMI that can be focused upon as future research ventures. 

\textit{Scalable Architecture:} A scalable and pervasive communication infrastructure is very much significant from the viewpoint of both construction and operation of a smart grid. A common feature of smart grid systems is that, a large amount of sensors are deployed over a wide area for implementing the complex monitoring and control functions.
Therefore, one prime challenge in smart grid is how to build a scalable
AMI communication architecture to handle the huge amount of data generated by those sensors. A smart grid communication infrastructure needs the scalability of accommodating more and more devices and services into it as well as more end-user interaction real-time monitoring of energy meters. It is evident that schemes based on conventional cryptographic operations are neither efficient nor scalable to the traffic density and resource constraints in a smart grid system. Therefore, we need scalable but secure and efficient schemes tailored specifically for smart grid AMI communications. This will enable secure and efficient processing of meter reading data collection and message distribution management. Very few works, such as~\cite{Zhou12}, investigated the scalability issue of distributed architecture in AMI. Also, key graph techniques can be used for addressing the scalability issue in AMI~\cite{Benmalek2016}. The proposed works considering scalable distributed architecture have shown promising outcomes with respect to key management of AMI. Therefore, future work should emphasize in building robust scalable distributed AMI communication architecture in smart grids.  

\textit{Content Centric Networking:} Content Centric Networking (CCN) is emerging as the fundamental paradigm for the future Internet and research initiatives are being undertaken for utilizing it further. The purpose of the CCN paradigm, is to shift the focus of communication from where information resides to what information is needed. By naming information at the networking layer, CCN favors the deployment of in-network caching and multicast mechanisms, thus, information will be targeted towards the interested hosts rather than hosts with specifically set destinations. The in-network caching of data in CCN improves network quality of services, especially, delivery latency. The Content Centric Networking (CCN) approach can be applied on AMI as a prospective future research area. Taking into consideration the caching advantage of CCN, it is widely believed that CCN can effectively reduce the AMI network bandwidth. So, CCN can play a major role in the traffic control development for the AMI system. The role of CCN in key management of AMI can be investigated to analyze how much effective role it plays.     

\textit{Defensive Mechanisms against Threats and Attacks:} AMI in smart grids are very much exposed to several threats and other vulnerabilities. AMI is an appropriate example of cyber physical systems comprising of different types of hardware, communication devices and MDMS. Therefore, AMI is very much exposed to cyber physical attacks. Smart meter networks and MDMS software must have sufficient security to prevent any unauthorized modification in software configurations, reading of recorded data, change in calibration, etc. Security in AMI depends on authentication mechanism, communication technologies and routing protocols. Another prevalent attack of AMI in smart grids is Denial of Service (DoS) attacks. These attacks are related with temporary or
permanent connect/disconnect AMI messages and communication link(wire-line or wireless) flooding/jamming by spoofing packets. In DoS attacks, an adversary may forge the demand request of a smart meter and keeps requesting a large amount of energy. Data integrity attacks also affect the normal operation of AMI by altering data timings, data analysis, false-data injection, data replay and data modifications. Thus, attacks that are very much prevalent in AMI need further exploration, specially with respect to key management and prospective solutions provided for defending such attacks. 

\textit{Consumer privacy and security:} Consumer privacy protection is a major requirement for present smart grid infrastructure. For this, smart meters should be equipped with latest storage components having high security precisions. It should not be accessed by unauthorized persons. Only authorized personnels must be able to decrypt the encrypted meter data. The privacy of customers and smart metering networks is important to the eventual acceptance by the public. One way to protect the consumers’ privacy is to make it impossible for unauthorized parties to distinguish load patterns and signatures~\cite{Kalogridis10}. Research in the area of providing consumer privacy and security is going on and smart meter users need to be reassured that their data is secure. Also, there is limited research on AMI authentication and confidentiality of user data privacy and behavior that needs serious attention. 

\textit{Efficiency:} From literature, it is quite evident that key management system is an essential component for secured AMI in smart grids. As smart grid contains millions of devices, spread across hundreds of organizations, the key management systems used must be scalable to high levels. Further, key management must offer strong security in terms of authentication and authorization, inter-organizational interoperability and the highest possible levels of efficiency to ensure that unnecessary cost due to overhead, provisioning and maintenance are minimized. It is likely that new and highly efficient key management systems (specialized to meet the requirements of smart grid) are needed.

\textit{Role of AMI in Smart Cities:} AMI could become one of the defining aspects of the smart cities of the future. Its widespread adoption could lead to a major impact in the efficient functioning of smart cities. The impact could be, unlimited savings and greater ease of use for consumers at all income levels and suppliers of utilities, by harnessing real-time data collection and consumer consumption patterns. If smart cities of the future depend on wireless networks to meet their utility demands, they can expect lower costs and greater bandwidth. Wireless networks also have the capacity to collect data from devices that are digitally-dormant today, as the existing IoT has shown that virtually everything can eventually be connected to the grid. Future works can be done where the role of key management system of AMI in smart cities are mainly focused upon.  

\textit{Standardization:} The establishment of industrial standards for smart metering techniques is an important part for the implementation of the future smart grid. Currently, extensive activities are being performed in standardizing components and communication between components of the advanced metering infrastructure. As standardization is an integral part for ensuring interworking of AMI equipment with heterogeneous manufactures, many regional and national institutes strive hard to achieve this goal. Nowadays, more and more Internet standards are applied in AMI. For example, the International Electro technical Commission (IEC) has defined application layer standards IEC 61968, that has created the common information models for AMI~\cite{Uslar}. The World Wide Web Consortium (W3C) committees has designed XML and EXI standards to address end-to-end messaging and formats. Further, the Institute of Electrical and Electronics Engineers (IEEE) 802.15.4 committee has addressed the gaps for cost-effective wireless mesh Neighbor Area Networks (NANs) at the physical and Media Access control (MAC) layers, etc. Thus, the standardization factor needs much more attention to make interoperability achievable for communication and information of AMI in smart grids.  

\section{CONCLUSION}
In this survey paper, we focus on the studies that investigate the challenges and opportunities of key management systems in AMI. We provide a comprehensive survey of the key management system of advanced metering infrastructure in smart grid. We first give a brief introduction of the smart grid and introduce the fundamental concepts of advanced metering infrastructure that have emerged with the smart grid. Further, we briefly describe how AMI is vulnerable to threats and defensive solutions can be provided by using key management systems. Next, we elaborate on the role of key management system in AMI followed by the different communication architectures that have adopted the key management systems of AMI. Then, we surveyed the state-of-the-art-works that developed mechanisms for efficient use of key management system in AMI. Security analysis of the schemes dealing with key management system in AMI is presented followed by performance analysis of those schemes with respect to storage, communication and computation overheads. Finally, potential research directions for key management system of AMI in smart grid security are identified. This key management system analysis of AMI brings new and promising perspectives and methodologies for future research in smart grid.

\bibliographystyle{IEEEtran}
\bibliography{Bibliography}

\begin{thebibliography}{10}
\providecommand{\url}[1]{#1}
\csname url@rmstyle\endcsname
\providecommand{\newblock}{\relax}
\providecommand{\bibinfo}[2]{#2}
\providecommand\BIBentrySTDinterwordspacing{\spaceskip=0pt\relax}
\providecommand\BIBentryALTinterwordstretchfactor{4}
\providecommand\BIBentryALTinterwordspacing{\spaceskip=\fontdimen2\font plus
\BIBentryALTinterwordstretchfactor\fontdimen3\font minus
  \fontdimen4\font\relax}
\providecommand\BIBforeignlanguage[2]{{%
\expandafter\ifx\csname l@#1\endcsname\relax
\typeout{** WARNING: IEEEtran.bst: No hyphenation pattern has been}%
\typeout{** loaded for the language `#1'. Using the pattern for}%
\typeout{** the default language instead.}%
\else
\language=\csname l@#1\endcsname
\fi
#2}}

\bibitem{Wu11}
D.~Wu and C.~Zhou, ``Fault-tolerant and scalable key management for smart
  grid,'' \emph{IEEE Transactions on Smart Grid}, vol.~2, no.~2, pp. 375--381,
  2011.

\bibitem{Pindoriya2013}
N.~M. Pindoriya, D.~Dasgupta, D.~Srinivasan, and M.~Carvalho, ``Infrastructure
  security for smart electric grids: A survey,'' in \emph{Optimization and
  Security Challenges in Smart Power Grids, Springer}, 2013, pp. 161--180.

\bibitem{Wang13}
W.~Wang and Z.~Lu, ``Cyber security in the smart grid: Survey and challenges,''
  \emph{Computer Networks}, vol.~57, no.~5, pp. 1344--1371, 2013.

\bibitem{Khasawneh17}
S.~Khasawneh and M.~Kadoch, ``Hybrid cryptography algorithm with precomputation
  for advanced metering infrastructure networks,'' \emph{Mobile Networks and
  Applications}, vol.~14, no.~4, pp. 1--12, 2017.

\bibitem{Yan12}
Y.~Yan, Y.~Qian, H.~Sharif, and D.~Tipper, ``A survey on cyber security for
  smart grid communications,'' \emph{IEEE Communications Surveys \& Tutorials},
  vol.~14, no.~4, pp. 998--1010, 2012.

\bibitem{Fang12}
X.~Fang, S.~Misra, G.~Xue, and D.~Yang, ``Smart grid— the new and improved
  power grid: A survey,'' \emph{IEEE Communications Surveys \& Tutorials},
  vol.~14, no.~4, pp. 944--980, 2012.

\bibitem{Deng17}
R.~Deng, G.~Xiao, R.~Lu, H.~Liang, and A.~V. Vasilakos, ``False data injection
  on state estimation in power systems—attacks, impacts, and defense: A
  survey,'' \emph{IEEE Transactions on Industrial Informatics}, vol.~13, no.~2,
  pp. 411--423, 2017.

\bibitem{Choi10}
D.~Choi, S.~Lee, D.~Won, and S.~Kim, ``Efficient secure group communications
  for {SCADA},'' \emph{IEEE Transactions on Power Delivery}, vol.~25, no.~2,
  pp. 714--722, 2010.

\bibitem{Erol-Kantarci15}
M.~E. Kantarci and H.~T. Mouftah, ``Energy-efficient information and
  communication infrastructures in the smart grid: A survey on interactions and
  open issues,'' \emph{IEEE Communications Surveys \& Tutorials}, vol.~17,
  no.~1, pp. 179--197, 2015.

\bibitem{YYan13}
Y.~Yan, Y.~Qian, H.~Sharif, and D.~Tipper, ``Survey on smart grid communication
  infrastructures: Motivations, requirements and challenges,'' \emph{IEEE
  Communications Surveys \& Tutorials}, vol.~15, no.~1, pp. 5--20, 2013.

\bibitem{Yan11}
Y.~Yan, Y.~Qian, and H.~Sharif, ``A secure and reliable in-network
  collaborative communication scheme for advanced metering infrastructure in
  smart grid,'' in \emph{Proc. of IEEE Wireless Communications and Networking
  Conference (IEEE WCNC)}, 2011, pp. 909--914.

\bibitem{Odelu16}
V.~Odelu, A.~K. Das, M.~Wazid, and M.~Conti, ``Provably secure authenticated
  key agreement scheme for smart grid,'' \emph{IEEE Transactions on Smart
  Grid}, vol.~27, no.~4, pp. 64--71, 2016.

\bibitem{US_report2}
DOE, ``Advanced metering infrastructure,'' {US} Department of Energy, Office of
  Electricity Delivery and Energy Reliability, Tech. Rep., 2008.

\bibitem{Mohassel14}
R.~R. Mohassel, A.~Fung, F.~Mohammadi, and K.~Raahemifar, ``A survey on
  advanced metering infrastructure,'' \emph{International Journal of Electrical
  Power \& Energy Systems}, vol.~63, pp. 473--484, 2014.

\bibitem{Kabalci16}
Y.~Kabalci, ``A survey on smart metering and smart grid communication,''
  \emph{Renewable and Sustainable Energy Reviews}, vol.~57, pp. 302--318, 2016.

\bibitem{Finster14}
S.~Finster and I.~Baumgart, ``Privacy-aware smart metering: A survey,''
  \emph{IEEE Communications Surveys \& Tutorials}, vol.~16, no.~3, pp.
  1732--1745, 2014.

\bibitem{Gungor11}
V.~C. Gungor, D.~Sahin, T.~Kocak, S.~Ergut, C.~Buccella, C.~Cecati, and G.~P.
  Hancke, ``Smart grid technologies: Communication technologies and
  standards,'' \emph{IEEE Transactions on Industrial Informatics}, vol.~7,
  no.~4, pp. 529--539, 2011.

\bibitem{Hasan2015}
M.~M. Hasan and H.~T. Mouftah, ``Encryption as a service for smart grid
  advanced metering infrastructure,'' in \emph{Proc. of IEEE Symposium on
  Computers and Communication (IEEE ISCC)}, 2015, pp. 216--221.

\bibitem{Seo2013}
S.~H. Seo, X.~Ding, and E.~Bertino, ``Encryption key management for secure
  communication in smart advanced metering infrastructures,'' in \emph{Proc. of
  IEEE International Conference on Smart Grid Communications (IEEE
  SmartGridComm)}, 2013, pp. 498--503.

\bibitem{Benmalek2016}
M.~Benmalek and Y.~Challal, ``{MK-AMI}: Efficient multi-group key management
  scheme for secure communications in {AMI} systems,'' in \emph{Proc. of IEEE
  Wireless Communications and Networking Conference (IEEE WCNC)}, 2016, pp.
  1--6.

\bibitem{Das12}
S.~Das, Y.~Ohba, M.~Kanda, D.~Famolari, and S.~K. Das, ``A key management
  framework for {AMI} networks in smart grid,'' \emph{IEEE Communications
  Magazine}, vol.~50, no.~8, 2012.

\bibitem{Kamto11}
J.~Kamto, L.~Qian, J.~Fuller, and J.~Attia, ``Light-weight key distribution and
  management for advanced metering infrastructure,'' \emph{Proc. of IEEE
  GLOBECOM Workshops}, pp. 1216--1220, 2011.

\bibitem{Rabieh17}
K.~Rabieh, M.~M. Mahmoud, K.~Akkaya, and S.~Tonyali, ``Scalable certificate
  revocation schemes for smart grid {AMI} networks using bloom filters,''
  \emph{IEEE Transactions on Dependable and Secure Computing}, vol.~14, no.~4,
  pp. 420--432, 2017.

\bibitem{Deng15}
R.~Deng, Z.~Yang, M.~Y. Chow, and J.~Chen, ``A survey on demand response in
  smart grids: Mathematical models and approaches,'' \emph{IEEE Transactions on
  Industrial Informatics}, vol.~11, no.~3, pp. 570--582, 2015.

\bibitem{Mohammadali18}
A.~Mohammadali, M.~S. Haghighi, M.~H. Tadayon, and A.~M. Nodooshan, ``A novel
  identity-based key establishment method for advanced metering infrastructure
  in smart grid,'' \emph{IEEE Transactions on Smart Grid}, pp. 1--9, 2018.

\bibitem{George16}
N.~George, S.~Nithin, and S.~K. Kottayil, ``Hybrid key management scheme for
  secure ami communications,'' in \emph{Proc. of 6th International Conference
  on Advances in Computing \& Communications (ICACC)}, vol.~93, 2016, pp.
  862--869.

\bibitem{Zhou12}
J.~Zhou, R.~Q. Hu, and Y.~Qian, ``Scalable distributed communication
  architectures to support advanced metering infrastructure in smart grid,''
  \emph{IEEE Transactions on Parallel and Distributed Systems}, vol.~23, no.~9,
  pp. 1632--1642, 2012.

\bibitem{Sharma15}
K.~Sharma and L.~M. Saini, ``Performance analysis of smart metering for smart
  grid: An overview,'' \emph{Renewable and Sustainable Energy Reviews},
  vol.~49, pp. 720--735, 2015.

\bibitem{Saputro11}
N.~Saputro, K.~Akkaya, and S.~Uludag, ``A survey of routing protocols for smart
  grid communications,'' \emph{Computer Networks}, vol.~56, no.~11, pp.
  2742--2771, 2011.

\bibitem{Colak16}
I.~Colak, S.~Sagiroglu, G.~Fulli, M.~Yesilbudak, and C.~Covrig, ``A survey on
  the critical issues in smart grid technologies,'' \emph{Renewable and
  Sustainable Energy Reviews}, vol.~54, pp. 396--405, 2016.

\bibitem{Sauter11}
T.~Sauter and M.~Lobashov, ``End-to-end communication architecture for smart
  grids,'' \emph{IEEE Transactions on Industrial Electronics}, vol.~58, no.~4,
  p. 1218–1228, 2011.

\bibitem{Wan14}
Z.~Wan, G.~Wang, Y.~Yang, and S.~Shi, ``\uppercase{SKM}: Scalable key
  management for advanced metering infrastructure in smart grids,'' \emph{IEEE
  Transactions on Industrial Electronics}, vol.~61, no.~12, pp. 7055--7066,
  2014.

\bibitem{Wang11}
W.~Wang, Y.~Xu, and M.~Khanna, ``A survey on the communication architectures in
  smart grid,'' \emph{Computer Networks}, vol.~55, no.~15, pp. 3604--3629,
  2011.

\bibitem{Jokar16}
P.~Jokar, N.~Arianpoo, and V.~Leung, ``A survey on security issues in smart
  grids,'' \emph{Security and Communication Networks}, vol.~9, no.~3, pp.
  262--273, 2016.

\bibitem{US_report1}
DOE, ``Communications requirements of smart grid technologies,'' {US}
  Department of Energy, pp. 1-69, Tech. Rep., 2010.

\bibitem{Taneja2013}
M.~Taneja, ``Lightweight security protocols for smart metering,'' in
  \emph{Proc. of IEEE Innovative Smart Grid Technologies Asia (ISGT Asia)},
  2013, pp. 1--5.

\bibitem{Kuzlu17}
M.~Kuzlu, M.~Pipattanasomporn, and S.~Rahman, ``Communication network
  requirements for major smart grid applications in {HAN}, {NAN} and {WAN},''
  \emph{Computer Networks}, vol.~67, pp. 74--88, 2017.

\bibitem{Chim15}
T.~W. Chim, S.~M. Yiu, V.~O. Li, L.~C. Hui, and J.~Zhong, ``{PRGA}:
  Privacy-preserving recording \& gateway-assisted authentication of power
  usage information for smart grid,'' \emph{IEEE Transactions on Dependable and
  Secure Computing}, vol.~12, no.~1, pp. 85--97, 2015.

\bibitem{Lunkeit013}
A.~Lunkeit, T.~Vo, and H.~Pohl, ``Threat modeling smart metering gateways,'' in
  \emph{Proc. of European Conference on Smart Objects, Systems and Technologies
  (SmartSysTech)}, 2013, pp. 1--5.

\bibitem{Sikora2013}
A.~Sikora, ``Implementation of standardized secure smart meter communication,''
  in \emph{Proc. of 35th International Telecommunications Energy Conference'
  Smart Power and Efficiency'(INTELEC)}, 2013, pp. 1--5.

\bibitem{Murrill12}
B.~J. Murrill, E.~C. Liu, and R.~M. Thompson, ``Smart meter data: Privacy and
  cyber security.'' Congressional Research Service, Library of Congress, Tech.
  Rep., 2012.

\bibitem{Molina-Markham10}
A.~M. Markham, P.~Shenoy, K.~Fu, E.~Cecchet, and D.~Irwin, ``Private memoirs of
  a smart meter,'' in \emph{Proc. of 2nd ACM Workshop on Embedded Sensing
  Systems for Energy-Efficiency in Building}, 2010, pp. 61--66.

\bibitem{Kalogridis10}
G.~Kalogridis, C.~Efthymiou, S.~Z. Denic, T.~A. Lewis, and R.~Cepeda, ``Privacy
  for smart meters: towards undetectable appliance load signatures,'' in
  \emph{Proc. of IEEE International Conference on Smart Grid Communications
  (IEEE SmartGridComm)}, 2010, pp. 232--237.

\bibitem{Yan13}
Y.~Yan, R.~Q. Hu, S.~K. Das, H.~Sharif, and Y.~Qian, ``An efficient security
  protocol for advanced metering infrastructure in smart grid,'' \emph{IEEE
  Network}, vol.~27, no.~4, pp. 64--71, 2013.

\bibitem{Anzalchi2015}
A.~Anzalchi and A.~Sarwat, ``A survey on security assessment of metering
  infrastructure in smart grid systems,'' in \emph{Proc. of IEEE SoutheastCon},
  2015, pp. 1--4.

\bibitem{Anas2012}
M.~Anas, A.~Javaid, N.and~Mahmood, S.~Raza, U.~Qasim, and Z.~Khan, ``Minimizing
  electricity theft using smart meters in {AMI},'' in \emph{Proc. of 7th IEEE
  International Conference on Parallel, Grid, Cloud and Internet Computing
  (3PGCIC)}, 2012, pp. 176--182.

\bibitem{Xia12}
J.~Xia and Y.~Wang, ``Secure key distribution for the smart grid,'' \emph{IEEE
  Transactions on Smart Grid}, vol.~3, no.~3, pp. 1437--1443, 2012.

\bibitem{Liu13}
N.~Liu, J.~Chen, L.~Zhu, J.~Zhang, and Y.~He, ``A key management scheme for
  secure communications of advanced metering infrastructure in smart grid,''
  \emph{IEEE Transactions on Industrial Electronics}, vol.~60, no.~10, pp.
  4746--4756, 2013.

\bibitem{Benmalek2015}
M.~Benmalek, Y.~Challal, and A.~Bouabdallah, ``Scalable multi-group key
  management for advanced metering infrastructure,'' in \emph{Proc. of IEEE
  International Conference on Computer and Information Technology (CIT)}, 2015,
  pp. 183--190.

\bibitem{Benmalek15}
M.~Benmalek and Y.~Challal, ``e{SKAMI}: Efficient and scalable multi-group key
  management for advanced metering infrastructure in smart grid,'' in
  \emph{Proc. of IEEE Wireless Communications and Networking Conference (IEEE
  WCNC)}, 2015, pp. 1--6.

\bibitem{Ben18}
M.~Benmalek, Y.~Challal, A.~Derhab, and A.~Bouabdallah, ``Ver{SAMI}: Versatile
  and scalable key management for smart grid {AMI} systems,'' \emph{Computer
  Networks}, vol. 132, pp. 161--179, 2018.

\bibitem{Yu15}
K.~Yu, M.~Arifuzzaman, Z.~Wen, D.~Zhang, and T.~Sato, ``A key management scheme
  for secure communications of information centric advanced metering
  infrastructure in smart grid,'' \emph{IEEE Transactions on Instrumentation
  and Measurement}, vol.~64, no.~8, pp. 2072--2085, 2015.

\bibitem{Wong00}
C.~K. Wong, M.~Gouda, and S.~S. Lam, ``Secure group communications using key
  graphs,'' \emph{IEEE/ACM Transactions on Networking}, vol.~8, no.~1, pp.
  16--30, 2000.

\bibitem{Parvez17}
I.~Parvez, A.~I. Sarwat, M.~T. Thai, and A.~K. Srivastava, ``A novel key
  management and data encryption method for metering infrastructure of smart
  grid,'' \emph{arXiv:1709.08505v1[cs.MA]}, 25 Sept. 2017.

\bibitem{Nicanfar14}
H.~Nicanfar, P.~Jokar, K.~Beznosov, and V.~C. Leung, ``Efficient authentication
  and key management mechanisms for smart grid communications,'' \emph{IEEE
  Systems Journal}, vol.~8, no.~2, pp. 629--640, 2014.

\bibitem{Nicanfar12}
H.~Nicanfar and V.~C. Leung, ``{EIBC}: Enhanced identity-based cryptography, a
  conceptual design,'' in \emph{Proc. of IEEE International Conference in
  Systems (SysCon)}, 2012, pp. 1--7.

\bibitem{Nabeel15}
M.~Nabeel, X.~Ding, S.~H. Seo, and E.~Bertino, ``Scalable end-to-end security
  for advanced metering infrastructures,'' \emph{Information Systems}, vol.~53,
  pp. 213--223, 2015.

\bibitem{Delavar17}
M.~Delavar, S.~Mirzakuchaki, M.~H. Ameri, and J.~Mohajeri, ``{PUF} based
  solutions for secure communications in advanced metering infrastructure
  {AMI},'' \emph{International Journal of Communication Systems}, vol.~67, pp.
  74--88, 2017.

\bibitem{Nabeel2012}
M.~Nabeel, S.~Kerr, X.~Ding, and E.~Bertino, ``Authentication and key
  management for advanced metering infrastructures utilizing physically
  unclonable functions,'' in \emph{Proc. of IEEE International Conference on
  Smart Grid Communications (IEEE SmartGridComm)}, 2012, pp. 324--329.

\bibitem{Zou08}
X.~Zou, Y.~S. Dai, and E.~Bertino, ``A practical and flexible key management
  mechanism for trusted collaborative computing,'' in \emph{Proc. of 27th IEEE
  Conference on Computer Communications (IEEE INFOCOM)}, 2008, pp. 538--546.

\bibitem{Shang10}
N.~Shang, M.~Nabeel, F.~Paci, and E.~Bertino, ``A privacy-preserving approach
  to policy-based content dissemination,'' in \emph{Proc. of 26th IEEE
  International Conference on Data Engineering (ICDE)}, March 2010, pp.
  944--955.

\bibitem{Seurin2012}
Y.~Seurin, ``On the exact security of schnorr-type signatures in the random
  oracle model,'' in \emph{Proc. of International Conference on the Theory and
  Applications of Cryptographic Techniques}, 2012, pp. 554--571.

\bibitem{Uslar}
M.~Uslar, M.~Specht, S.~Rohjans, J.~Trefke, and J.~M. González, \emph{The
  Common Information Model CIM: IEC 61968/61970 and 62325-A practical
  introduction to the CIM}.\hskip 1em plus 0.5em minus 0.4em\relax Springer
  Science \& Business Media, Springer, 2012.

\end{thebibliography}

\end{document}